\documentclass{naturedraft}
\usepackage{color, graphicx, amssymb}
\newcommand{\Ket}[1]{\vert \! #1 \, \rangle}

\title{Long coherence of electron spins coupled to a nuclear spin bath}

\author{Hendrik Bluhm$^{1*}$, Sandra Foletti$^{1*}$, Izhar Neder$^{1}$, 
Mark Rudner$^{1}$, Diana Mahalu$^2$, Vladimir Umansky$^2$ \& Amir Yacoby$^1$}

\begin{document}
\maketitle

\begin{affiliations}
\item Department of Physics, Harvard University, Cambridge, MA 02138, USA
\item Braun Center for Submicron Research, Department of Condensed Matter Physics, Weizmann Institute of Science, Rehovot 76100, Israel
\item [*] These authors contributed equally to this work.
\end{affiliations}

\begin{abstract}
Qubits, the quantum mechanical bits required for quantum computing,
must retain their fragile quantum states over long periods of time.
In many types of electron spin qubits, the primary source of decoherence
is the interaction between the electron spins and  
nuclear spins of the host lattice. 
For electrons in gate defined GaAs quantum dots, previous spin echo
measurements have revealed coherence times of about 1 $\mu$s 
at low magnetic fields below 100 mT (ref \citen{Petta2005, Koppens2008}).
Here, we show that coherence in such devices can actually survive to
much longer times, and provide a detailed understanding of 
the measured nuclear spin induced decoherence.
At fields above a few hundred millitesla, the 
coherence time measured using a
single-pulse spin echo extends to 30 $\mu$s.  
At lower magnetic
fields, the echo first collapses, but then revives at later times
given by the period of the relative Larmor precession of different
nuclear species.
This behavior was recently predicted\cite{Cywinski2009, Cywinski2009B}, and as we show can be quantitatively accounted for 
by a semi-classical model for the electron spin dynamics in the 
presence of a nuclear spin bath. 
Using a multiple-pulse Carr-Purcell-Meiboom-Gill echo sequence, 
the decoherence time can be extended to more than 200 $\mu$s,
which represents an improvement by two orders of magnitude compared to
previous measurements\cite{Petta2005, Koppens2008, Greilich2006}.
This demonstration of effective methods to mitigate 
nuclear spin induced decoherence 
puts the quantum error correction threshold within reach.
\end{abstract}

The  promise of quantum dot spin qubits as a solid state
approach to quantum computing is demonstrated by the successful realization of
initialization, control and single shot readout of electron spin
qubits in GaAs quantum dots using optical\cite{Press2008},
magnetic\cite{Koppens2005}, and fully electrical\cite{Foletti2009,
Nowack2007, Barthel2009} techniques. 
To further advance spin based quantum computing, it is vital 
to mitigate decoherence due to the interaction of the
electron spin with the spins of nuclei of the host material.
Understanding the dynamics of this system is also of great fundamental 
interest\cite{Coish2008,Chen:semiclass}.

Through the hyperfine interaction, an electron spin in a GaAs quantum dot
is subjected to an effective magnetic field produced by the nuclear spins.
Under typical experimental conditions, this so-called ``Overhauser
field'' has a random magnitude and direction.
Typically, measurements of the coherent electron spin precession involve
averaging over many experimental runs, and thus over many Overhauser
field configurations.
As a result, the coherence signal is suppressed for evolution times
$\tau \gtrsim T_2^* \approx 10$ ns (ref \citen{Petta2005}).
However, the nuclear spins evolve much more slowly than
the electron spins, so that the Overhauser field is nearly static 
for a sufficiently short duration of electron spin evolution.
Therefore, it is possible to partially eliminate the effect of the
random nuclear field by flipping the electron spin halfway
though an interval of free precession\cite{Hahn1950}, a procedure known
as Hahn-echo.
The  random contributions of the Overhauser field to the 
electron spin precession before and after the
spin-reversal then approximately cancel out.
For longer evolution times, the effective field acting on the electron
spin generally changes between the two halves of the precession
interval.
This change leads to an eventual loss of coherence on a time scale 
determined by the details of the nuclear spin dynamics.

Previous Hahn-echo experiments in GaAs quantum dot spin qubits have
demonstrated spin dephasing times of around 1 $\mu$s
at relatively low magnetic fields up to 100 mT (ref \citen{Petta2005,
Koppens2008}). 
Recent theoretical studies of decoherence due to the hyperfine
interaction\cite{Yao2006, Cywinski2009,Cywinski2009B} are generally
consistent with these experimental results, but predict revivals of
the echo signal after several microseconds, as also seen in other
systems\cite{Childress2006}.  This prediction already indicates that
the initial decay of the echo does not reflect irreversible
decoherence, but is a consequence of the coherent Larmor precession of
the nuclei.  Theoretical work also predicted much longer coherence
times at higher external magnetic fields\cite{Witzel2006} or when 
using more advanced pulse sequences\cite{Witzel:MultPulse,
Lee2008}. 
The classic example is the Carr-Purcell-Meiboom-Gill (CPMG) 
sequence\cite{Meiboom1958,Petta2005}, but several alternatives 
have recently been developed\cite{Khodjasteh:CDD,Uhrig2007} and 
demonstrated\cite{Biercuk:ODD, Du:ODD}.
The performance of such schemes is expected to improve as more control 
pulses are added\cite{Lee2008}.  
Here, we provide direct experimental confirmations for all the above 
predictions. 

\begin{figure}
\begin{center}
\includegraphics[width=9cm]{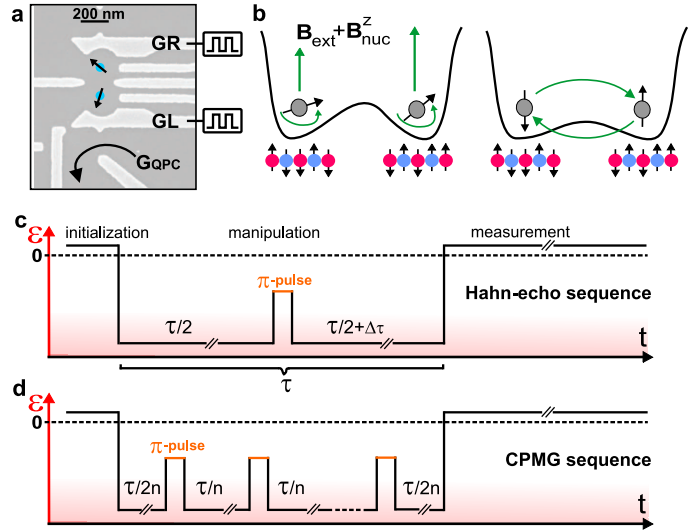}
\end{center}
\caption{\textbf{Qubit control. }\textbf{a,} SEM
  micrograph of a device similar to the one used. Metal gates (bright
  structures) are negatively biased  to confine two
  electrons. The charge state of the double quantum dot is determined
  by measuring the conductance through the capacitively coupled
  quantum point contact, $G_{QPC}$. The separation
  between the two electrons is controlled with nanosecond
  time resolution via the voltages on GR and GL.
  \textbf{b,} Left: an initially prepared
  singlet state oscillates between $S$ and $T_0$ with frequency 
  $g^* \mu_{B}\Delta B_{\rm nuc}^z/\hbar$, which changes over time due to
  slow fluctuations of the hyperfine field gradient $\Delta B_{\rm nuc}^z$.
  Right: switching on the tunnel coupling between the two dots
  leads to the coherent exchange of the electron spins.
  \textbf{c,} Hahn-echo sequence: after evolving
  for a time $\tau/2$, the two electrons are exchanged with a
  $\pi$-pulse. The singlet state is recovered after further evolution
  for another $\tau/2$, independent of
  $\Delta B_{\rm nuc}^z$.  \textbf{d,} Carr-Purcell-Meiboom-Gill sequence:
  in this higher order decoupling sequence, $n$ $\pi$-pulses at
  time intervals $\tau/n$ are applied.}
\label{fig1}
\end{figure}

The spin qubit studied in this work consists of two isolated electrons
confined in a double quantum dot, created by applying negative
voltages to metallic gates that locally deplete a two dimensional
electron gas (2DEG) 90 nm below the wafer surface (see
Fig. \ref{fig1}a). 
The Hilbert space of our logical qubit is spanned
by the states $\Ket{\uparrow \downarrow}$ and $\Ket{\downarrow
\uparrow}$, i.e. the $m = 0$ subspace of two separated spins.
The arrows represent the alignment of the electron spins in each of
the dots relative to an external magnetic field, $B_{ext}$, which is
oriented in the plane of the 2DEG.
The remaining two states, $T_+ \equiv \Ket{\uparrow \uparrow}$ and
$T_- \equiv \Ket{\downarrow \downarrow}$, are energetically separated
by the Zeeman energy $E_Z = g^* \mu_B B_{ext}$ ($g^* = -0.44$ is the
$g$-factor in GaAs) and are not used for information storage.
Tunnel coupling to the leads is used for initialization, while 
inter-dot tunnel coupling allows spin exchange between the
dots. This exchange interaction is  modulated via the detuning
$\varepsilon$, which is the difference between the electrostatic
potentials in the two dots.
This parameter is controlled via rapid,
antisymmetric changes of the voltages on gates GL and GR (see
Fig. \ref{fig1}) applied via high frequency coaxial lines,
which enables fast electrical control of the qubit\cite{Levy2002,
Petta2005, Foletti2009}

The experimental procedures follow those of ref \citen{Petta2005}.
We initialize the system at a large detuning, 
where the ground state is a spin singlet with both electrons residing
in a single dot.
As $\varepsilon$ is swept to negative values, the electrons separate
into different dots, thus preparing the singlet state $S \equiv
(\Ket{\uparrow \downarrow} - \Ket{\downarrow \uparrow})/\sqrt{2}$.
For very large negative detunings, 
the electron spins in the two dots are decoupled, and each
individually experiences a Zeeman field composed of the homogeneous
external field and a fluctuating local hyperfine field.
 A difference $\Delta B_{\rm nuc}^z$ between the $z$-components of the
hyperfine fields in the two dots leads to an energy splitting between
the basis states $\Ket{\uparrow \downarrow}$ and $\Ket{\downarrow
\uparrow}$.
This splitting causes precession between the singlet $S$
and the triplet $T_0 \equiv (\Ket{\uparrow \downarrow} +
\Ket{\downarrow \uparrow})/\sqrt{2}$, and its fluctuations lead to dephasing of the qubit.
We implement the echo $\pi$-pulses by pulsing to small negative detunings,
where inter-dot tunneling leads to an exchange
splitting between $S$ and $T_0$. This splitting drives coherent
oscillations between the states $\Ket{\uparrow \downarrow}$ and
$\Ket{\downarrow \uparrow}$. 
The pulse profiles for the Hahn-echo and CPMG sequence are shown in Fig.
\ref{fig1}c,d.

Readout of the final qubit state is accomplished by switching to positive
detuning, $\varepsilon > 0$, where the state with both electrons
sitting in the same dot is preferred for the spin singlet, but is
energetically excluded for a spin-triplet due to the Pauli exclusion
principle.
The two spin states thus acquire different charge densities.
To sense this difference, we use a proximal quantum point contact
(QPC), whose conductance depends on the local electrostatic
environment\cite{Field1993}.
After averaging over many identical pulse cycles, the mean QPC
conductance, $G_{QPC}$, reflects the probability to find the qubit
in the singlet state at the end of each cycle.
The echo amplitudes presented below are
normalized such that they are unity at short 
times (no decoherence) and eventually drop to zero 
for a fully randomized state. This normalization 
eliminates $\tau$-independent contrast 
losses (see Supplementary Material).
Fig. \ref{fig2}a shows the Hahn-echo signals for different
magnetic fields $B_{ext}$.

\begin{figure}
\begin{center}
\includegraphics[width=9cm]{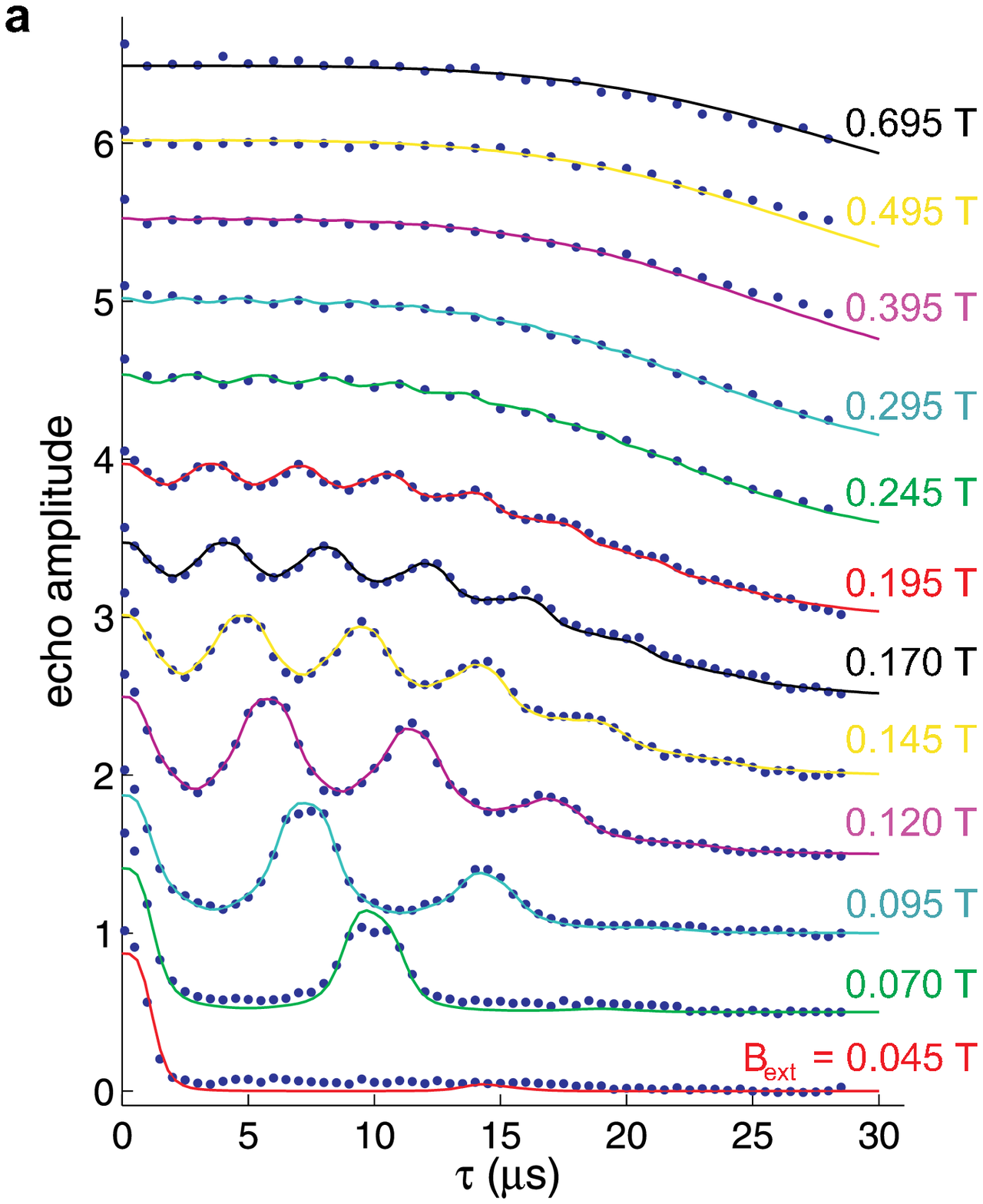}\qquad
\includegraphics[width=6.5cm]{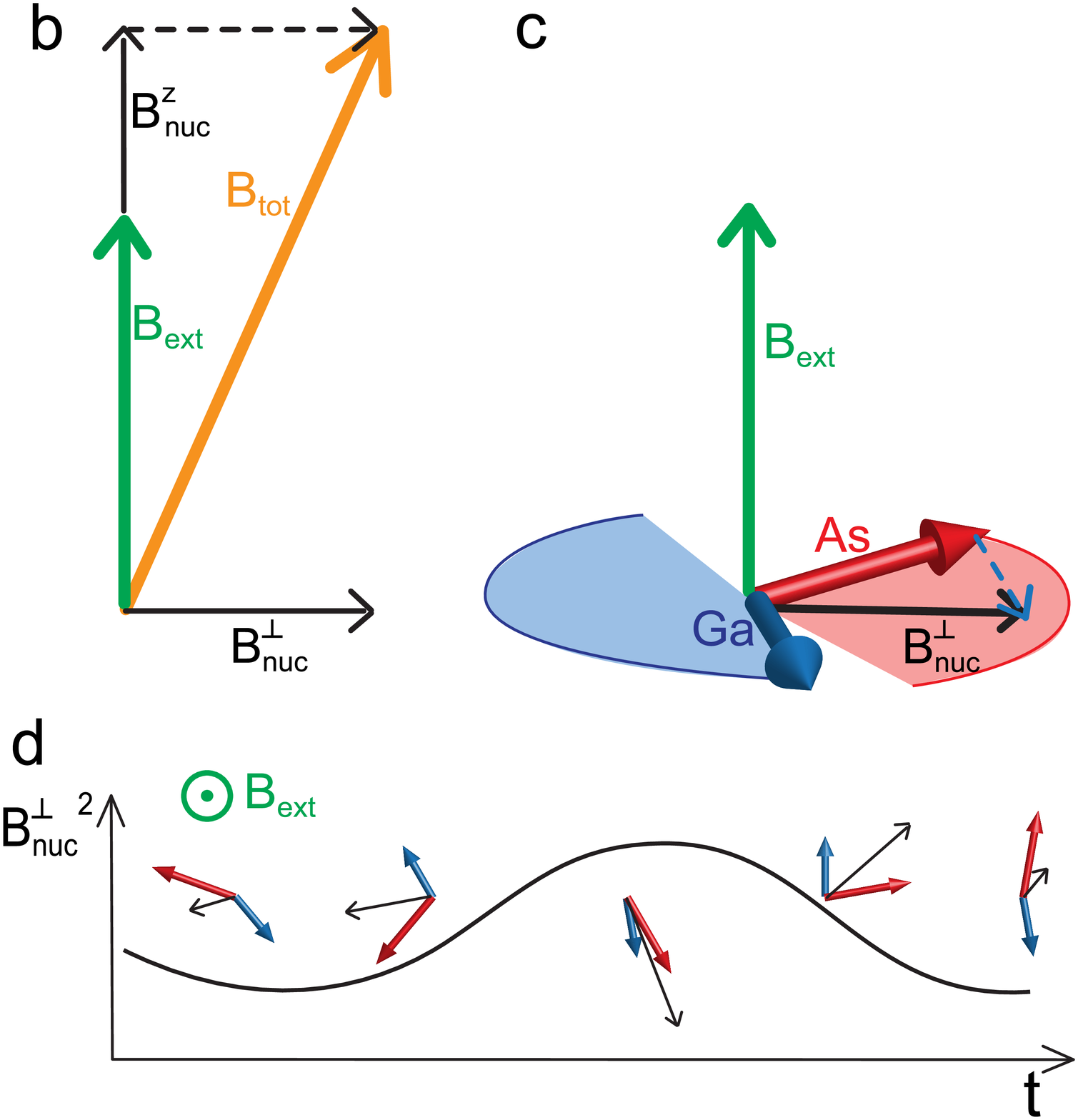}
\end{center}
\caption{\textbf{Echo amplitude.}
{\bf a,} Echo signal as a function of the total evolution time, $\tau$, 
for different values of magnetic field. The fits to the data are obtained
by extending the model of ref \citen{Cywinski2009} to include a spread 
$\delta B_{\rm loc}$ of the nuclear Larmor frequencies\cite{Neder2010} and multiplying 
with $\exp((-\tau/T_{SD})^4)$. Curves are offset for
clarity and normalized as discussed in the Supplementary Material.
{\bf b,} The total Zeeman field seen by the electron is the vector sum of 
the external field and the Overhauser fields parallel and perpendicular 
to it. {\bf c,} The three nuclear species (only two shown for clarity) 
contributing to the  Overhauser field precess at different Larmor 
frequencies in the external field. {\bf d,} As a result of the relative
precession (top), the 
total transverse nuclear field oscillates at the Larmor 
frequency difference(s) (bottom).
}
\label{fig2}
\end{figure}

At high fields we find a monotonic decay of the Hahn-echo signal with
the total duration of free precession, $\tau$.  The echo initially
decays very slowly (approximately proportional to $\tau^4$), and
 is suppressed to  $1/e$ after 30 $\mu$s.
As  the magnetic field is reduced, the echo signal develops
oscillations with a time scale of microseconds.  For even lower fields,
the oscillations evolve into full collapses of the signal, with
revivals at later times on a ten microsecond time scale.  These
revivals were predicted in refs \citen{Cywinski2009, Cywinski2009B}
based on a quantum mechanical treatment of the hyperfine interaction
between electron and nuclear spins.  Below we outline a semiclassical
model\cite{Neder2010} that can reproduce the lowest-order result 
of refs \citen{Cywinski2009, Cywinski2009B} and accounts for
additional effects that are essential for fitting our data.
This model provides the theoretical echo signal,
$C(\tau) \equiv  2 p(S)- 1
 = -\mathrm{Re} \langle\uparrow\downarrow|\rho(\tau)\Ket{\downarrow\uparrow}$, 
where $p(S)$ is the probability of finding 
the electron in a singlet state and 
$\rho(\tau)$ is the qubit's density matrix at the end of the 
evolution time. We have used this model to 
produce the quantitatively accurate fits displayed in Fig. \ref{fig2}a.

For each electron spin, the Zeeman energy splitting is proportional to
the total magnetic field 
$B_{tot} = \sqrt{(B_{ext} + B_{\rm nuc}^z)^2 + {B_{\rm nuc}^\perp}^2} \approx
B_{ext} + B_{\rm nuc}^z + {B_{\rm nuc}^\perp}^2/2 B_{ext}$ (Fig. \ref{fig2}b).  
Time-dependence of the parallel and transverse nuclear components,
$B_{\rm nuc}^z$ and ${B_{\rm nuc}^\perp}^2$,
can lead to dephasing of the electron spin.  Assuming statistical
independence between $B_{\rm nuc}^z$ and ${B_{\rm nuc}^\perp}^2$, 
the theoretical echo signal can be written as a product, $C(\tau) =  A_z(\tau)
R_\perp(\tau)$, where $A_z(\tau)$ and $R_\perp(\tau)$ account for the
contributions of $B_{\rm nuc}^z$ and ${B_{\rm nuc}^\perp}$ to the electron spin
precession.  
In the experimental range of the magnetic
fields, the time-dependence of $B_{\rm nuc}^z$ is mainly caused by spectral
diffusion due to the dipole-dipole interaction between nuclear
spins.  This process is predicted to 
lead to a decay of the
form $A_z(\tau) = \exp(-(\tau/T_{SD})^4)$ (ref \citen{Witzel2006, Yao2006}).  As we will now discuss,
$R_\perp(\tau)$ (see Supplementary Eq. 1)
has a more interesting non-monotonic structure which
arises from the relative precession of nuclear spins in the external field 
with different Larmor frequencies.

The transverse hyperfine field, $\vec B_{\rm nuc}^\perp$, is a vector sum of
contributions from the three nuclear species $^{69}$Ga, $^{71}$Ga and
$^{75}$As.  Due to the different precession rates of these species 
(Fig. \ref{fig2}c),
$B_{\rm nuc}^\perp(t)^2$ contains harmonics at the three relative Larmor
frequencies (Fig. \ref{fig2}d) in addition to a constant term.  The
contribution of the constant term to the singlet return probability is
eliminated by the echo pulse.  For a general free precession period,
the time-dependence leads to a suppression of the echo signal.
However, if the precession interval $\tau/2$ is a multiple of all
three relative Larmor periods, the oscillatory components contribute
no net phase to the electron spin evolution.  As a result, the echo
amplitude revives whenever the commensurability condition is met.
Averaging the singlet return probability  
over initial Overhauser field configurations\cite{Neder2010} leads to the 
collapse-and-revival behavior predicted in refs 
\citen{Cywinski2009, Cywinski2009B}.

At low fields, the echo envelope decays more quickly than at high
fields (see Fig. \ref{fig2}a).  This field dependence can be accounted
for by including a spread of the Larmor precession frequencies for
each nuclear species.  Such a variation is also manifest in the width
of NMR lines and naturally arises from dipolar and other interactions
between nuclei\cite{Sundfors:NMR}.  We model it as a shift of the
magnetic field acting on each individual nuclear spin by an amount
$B_{\rm loc}$, where $B_{\rm loc}$ is a Gaussian random variable with
standard deviation $\delta B_{\rm loc}$.  This spread of precession
frequencies leads to an aperiodic time-dependence of ${B_{\rm
nuc}^\perp}^2$, which cannot be removed by the electron spin echo.

Using the above model (see also Supplementary Eq. 1),
we have fit all the data in Fig. \ref{fig2}a
with a single set of field-independent parameters which were chosen to 
obtain a good match with all datasets: the number of nuclei 
in each of the two dots, $N$, the spectral diffusion time constant, 
$T_{SD}$, and $\delta B_{loc}$. In addition, the scale factor for each 
dataset was allowed 
to vary to account for the imperfect normalization of the data.
The value of $N$ determines the depths of the dips between revivals.  The
best fit yields $N = 4.4 \times 10^6$, which is in good agreement with
an independent determination from a measurement of $T_2^*
= \sqrt{N} \hbar/g^* \mu_B \cdot 4.0$ T giving $N = 4.9 \times
10^6$ (see ref \citen{Taylor2007}, Supplementary Material).  From the fit
we also obtain $T_{SD} \approx 37 \mu$s and $\delta B_{\rm loc}$ = 0.3
mT.  The measured NMR line width in pure GaAs is about 0.1 mT
(ref \citen{Sundfors:NMR}).  A possible origin for the larger field
inhomogeneity found here is the quadrupole splitting arising from the
presence of the two localized electrons\cite{Hester:defect}.  The
inhomogeneity of the Knight shift is expected to have a similar but quantitatively
negligible effect for our parameters.  The value of $T_{SD}$ is
consistent with theoretical estimates (see Supplementary Material and
ref \citen{Witzel2006}).  Interestingly, the spread of nuclear Larmor
frequencies, captured by $\delta B_{\rm loc}$, contributes
significantly to the echo decay even at the highest fields
investigated.  We have also verified that the Hahn-echo lifetime is
not significantly affected by dynamic nuclear polarization, which can
be used to increase $T_2^*$ (ref \citen{Bluhm2009}, Supplementary Material).

In order to measure the long Hahn-echo decay times of up to 30 $\mu$s,
it was necessary to systematically optimize the pulses (see
Supplementary Material).  Small differences in the gate voltages
before and after the $\pi$-pulse shift the electronic wave function
relative to the inhomogeneously polarized nuclei. 
Such shifts cause the electrons to sample
different Overhauser fields at different times, and thus lead to an
imperfect echo. We have minimized this effect by compensating for a
systematic drift of $\varepsilon$ over the course of each pulse
sequence (see Supplementary Material).

\begin{figure}
\begin{center}
\includegraphics[width=9cm]{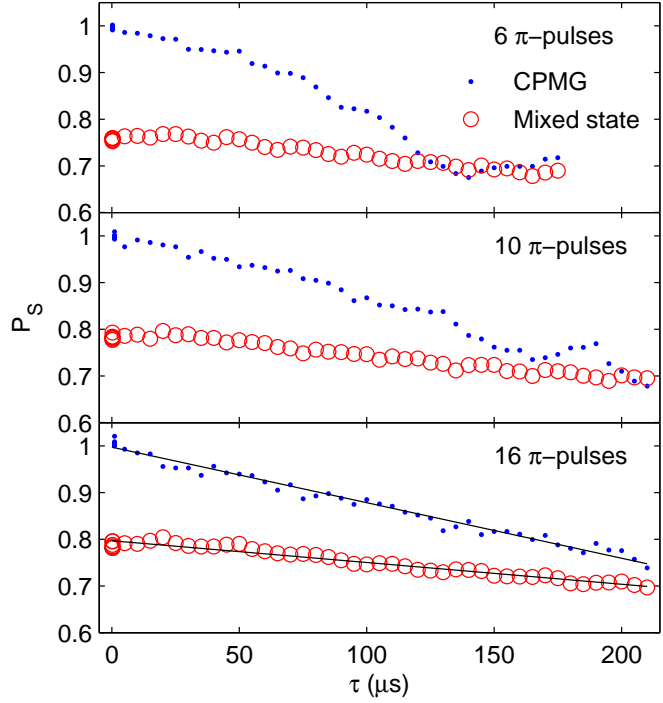}
\end{center}
\caption{\textbf{CPMG decoupling experiments with 6, 10 and 16 $\pi$-pulses}
at $B_{ext}$ = 0.4 T.
The blue dots show the readout signal of the CPMG pulses, the red
circles represent reference measurements with the same evolution time
without any $\pi$-pulses (equivalent to $T_2^*$ measurements), which
produce a completely dephased state. $P_S$ is the sensor signal normalized 
by the DC contrast associated with the transfer of an electron from one 
dot to the other, so that a singlet
corresponds to $P_S = 1$ (See Supplementary Material). 
Inelastic decay during the readout phase and
possibly other visibility loss mechanisms increase $P_S$ for the mixed
state above the ideal value of 0.5.  The linear trends in the
reference and the initial decay of the CPMG signal likely reflect
leakage out of the logical subspace.  The linear fits to the 16-pulse
data (black lines) intersect at $\tau$ = 276 $\mu$s, which can be
taken as a rough estimate or lower bound of the coherence time.
}
\label{fig4}
\end{figure}

Substantially longer coherence times are expected for more elaborate
decoupling sequences\cite{Lee2008}.
We implemented the CPMG sequence\cite{Meiboom1958}, 
which consists of an $n$-fold repetition of the
Hahn-echo, thus requiring $n$ $\pi$-pulses, as shown in
Fig. \ref{fig1}d.  Fig. \ref{fig4} shows data for $n$ = 6, 10 and
16. For $n$ = 16, the echo signal clearly persists for more than 200
$\mu$s. The field dependence for $n$ = 4 is reported in the
Supplementary Material.  To verify the interpretation of the data, we
have measured the dependence of the echo on small changes in the final 
free precession time
and the duration of the exchange pulses for $n$ = 10,
$\tau$ = 5 and 120 $\mu$s (Supplementary Material).  Because of
the large number of potential tuning parameters, we have not optimized
these CPMG pulses. We expect that with improved pulses, the same
extension of the coherence time could be achieved with fewer pulses.
The linear initial decay of the signal in Fig. \ref{fig4} is not well
understood.  The similar variation of the reference signal
corresponding to a completely mixed state is suggestive of a single-electron $T_1$ process causing
leakage into the $T^+$ and $T^-$ states (see Supplementary Material).
The decay time constant 
sets a lower bound for the largest achievable coherence time.

Our measurements demonstrate coherence times of GaAs spin qubits 
of at least 200 $\mu$s, two orders of magnitude larger than previously shown. 
The duration of each of the $\pi$-pulses could easily be reduced 
below the 6 ns used here. Thus, more than $10^5$ operations
could be carried out within the coherence time, well in excess of the
commonly quoted error correction threshold of $\sim 10^4$.
Furthermore, one may hope to achieve millisecond scale coherence times with 
improved decoupling sequences\cite{Lee2008} without adding complexity.
The excellent agreement with the model for the field and time dependence
of the Hahn-echo revivals shows that many aspects of
the dephasing of electron spins due to the nuclear hyperfine
interaction are now well understood.
The insight gained may also help pave the way towards probing macroscopic
quantum effects in a mesoscopic ensemble of a few million nuclear spins.

\bibliography{../../bibdata}

\begin{thebibliography}{10}
\expandafter\ifx\csname url\endcsname\relax
  \def\url#1{\texttt{#1}}\fi
\expandafter\ifx\csname urlprefix\endcsname\relax\def\urlprefix{URL }\fi
\providecommand{\bibinfo}[2]{#2}
\providecommand{\eprint}[2][]{\url{#2}}

\bibitem{Petta2005}
\bibinfo{author}{Petta, J.~R.} \emph{et~al.}
\newblock \bibinfo{title}{Coherent manipulation of coupled electron spins in
  semiconductor quantum dots}.
\newblock \emph{\bibinfo{journal}{Science}} \textbf{\bibinfo{volume}{309}},
  \bibinfo{pages}{2180} (\bibinfo{year}{2005}).

\bibitem{Koppens2008}
\bibinfo{author}{Koppens, F. H.~L.}, \bibinfo{author}{Nowack, K.} \&
  \bibinfo{author}{Vandersypen, L. M.~K.}
\newblock \bibinfo{title}{Spin echo of a single spin in a quantum dot}.
\newblock \emph{\bibinfo{journal}{Phys. Rev. Lett.}}
  \textbf{\bibinfo{volume}{100}}, \bibinfo{pages}{236802}
  (\bibinfo{year}{2008}).

\bibitem{Cywinski2009}
\bibinfo{author}{Cywinski, L.}, \bibinfo{author}{Witzel, W.~M.} \&
  \bibinfo{author}{{Das Sarma}, S.}
\newblock \bibinfo{title}{Pure quantum dephasing of a solid-state electron spin
  qubit in a large nuclear spin bath coupled by long-range hyperfine mediated
  interactions}.
\newblock \emph{\bibinfo{journal}{Phys. Rev. B}} \textbf{\bibinfo{volume}{79}},
  \bibinfo{pages}{245314} (\bibinfo{year}{2009}).

\bibitem{Cywinski2009B}
\bibinfo{author}{Cywinski, L.}, \bibinfo{author}{Witzel, W.~M.} \&
  \bibinfo{author}{{Das Sarma}, S.}
\newblock \bibinfo{title}{Electron spin dephasing due to hyperfine interactions
  with a nuclear spin bath}.
\newblock \emph{\bibinfo{journal}{Phys. Rev. Lett.}}
  \textbf{\bibinfo{volume}{102}}, \bibinfo{pages}{057601}
  (\bibinfo{year}{2009}).

\bibitem{Greilich2006}
\bibinfo{author}{Greilich, A.} \emph{et~al.}
\newblock \bibinfo{title}{Mode locking of electrons spin coherence in singly
  charged quantum dots}.
\newblock \emph{\bibinfo{journal}{Science}} \textbf{\bibinfo{volume}{313}},
  \bibinfo{pages}{341} (\bibinfo{year}{2006}).

\bibitem{Press2008}
\bibinfo{author}{Press, D.}, \bibinfo{author}{Ladd, T.},
  \bibinfo{author}{Zhang, B.} \& \bibinfo{author}{Yamamoto, Y.}
\newblock \bibinfo{title}{Complete quantum control of a single quantum dot spin
  using ultrafast optical pulses}.
\newblock \emph{\bibinfo{journal}{Nature}} \textbf{\bibinfo{volume}{456}},
  \bibinfo{pages}{218} (\bibinfo{year}{2008}).

\bibitem{Koppens2005}
\bibinfo{author}{{Koppens}, F. H.~L.} \emph{et~al.}
\newblock \bibinfo{title}{Control and detection of singlet-triplet mixing in a
  random nuclear field}.
\newblock \emph{\bibinfo{journal}{Science}} \textbf{\bibinfo{volume}{309}},
  \bibinfo{pages}{1346} (\bibinfo{year}{2005}).

\bibitem{Foletti2009}
\bibinfo{author}{Foletti, S.}, \bibinfo{author}{Bluhm, H.},
  \bibinfo{author}{Mahalu, D.}, \bibinfo{author}{Umansky, V.} \&
  \bibinfo{author}{Yacoby, A.}
\newblock \bibinfo{title}{Universal quantum control in two-electron spin
  quantum bits using dynamic nuclear polarization.}
\newblock \emph{\bibinfo{journal}{Nature Physics}}
  \textbf{\bibinfo{volume}{5}}, \bibinfo{pages}{903} (\bibinfo{year}{2009}).

\bibitem{Nowack2007}
\bibinfo{author}{Nowack, K.~C.}, \bibinfo{author}{Koppens, F. H.~L.},
  \bibinfo{author}{Nazarov, Y.~V.} \& \bibinfo{author}{Vandersypen, L. M.~K.}
\newblock \bibinfo{title}{Coherent control of a single electron spin with
  electric fields}.
\newblock \emph{\bibinfo{journal}{Science}} \textbf{\bibinfo{volume}{318}},
  \bibinfo{pages}{1430} (\bibinfo{year}{2007}).

\bibitem{Barthel2009}
\bibinfo{author}{Barthel, C.}, \bibinfo{author}{Reilly, D.~J.},
  \bibinfo{author}{Marcus, C.~M.}, \bibinfo{author}{Hanson, M.~P.} \&
  \bibinfo{author}{Gossard, A.~C.}
\newblock \bibinfo{title}{Rapid single-shot measurement of a singlet-triplet
  qubit}.
\newblock \emph{\bibinfo{journal}{Phys. Rev. Lett.}}
  \textbf{\bibinfo{volume}{103}}, \bibinfo{pages}{160503}
  (\bibinfo{year}{2009}).

\bibitem{Coish2008}
\bibinfo{author}{Coish, W.~A.}, \bibinfo{author}{Fischer, J.} \&
  \bibinfo{author}{Loss, D.}
\newblock \bibinfo{title}{Exponential decay in a spin bath}.
\newblock \emph{\bibinfo{journal}{Phys.\ Rev.\ B.}}
  \textbf{\bibinfo{volume}{77}}, \bibinfo{pages}{125329}
  (\bibinfo{year}{2008}).

\bibitem{Chen:semiclass}
\bibinfo{author}{Chen, G.}, \bibinfo{author}{Bergman, D.~L.} \&
  \bibinfo{author}{Balents, L.}
\newblock \bibinfo{title}{Semiclassical dynamics and long-time asymptotics of
  the central-spin problem in a quantum dot}.
\newblock \emph{\bibinfo{journal}{Phys. Rev. B}} \textbf{\bibinfo{volume}{76}},
  \bibinfo{pages}{045312} (\bibinfo{year}{2007}).

\bibitem{Hahn1950}
\bibinfo{author}{Hahn, E.~L.}
\newblock \bibinfo{title}{Spin echoes}.
\newblock \emph{\bibinfo{journal}{Phys. Rev.}} \textbf{\bibinfo{volume}{80}},
  \bibinfo{pages}{580} (\bibinfo{year}{1950}).

\bibitem{Yao2006}
\bibinfo{author}{Yao, W.}, \bibinfo{author}{Liu, R.~B.} \&
  \bibinfo{author}{Sham, L.~J.}
\newblock \bibinfo{title}{Theory of electron spin decoherence by interacting
  nuclear spins in a quantum dot}.
\newblock \emph{\bibinfo{journal}{Phys.\ Rev.\ B.}}
  \textbf{\bibinfo{volume}{74}}, \bibinfo{pages}{195301}
  (\bibinfo{year}{2006}).

\bibitem{Childress2006}
\bibinfo{author}{Childress, L.} \emph{et~al.}
\newblock \bibinfo{title}{Coherent dynamics of coupled electron and nuclear
  spin qubits in diamond}.
\newblock \emph{\bibinfo{journal}{Science}} \textbf{\bibinfo{volume}{314}},
  \bibinfo{pages}{281} (\bibinfo{year}{2006}).

\bibitem{Witzel2006}
\bibinfo{author}{Witzel, W.~M.} \& \bibinfo{author}{Sarma, S.~D.}
\newblock \bibinfo{title}{Quantum theory for electron spin decoherence induced
  by nuclear spin dynamics in semiconductor quantum computer architecture}.
\newblock \emph{\bibinfo{journal}{Phys.\ Rev.\ B.}}
  \textbf{\bibinfo{volume}{74}}, \bibinfo{pages}{035322}
  (\bibinfo{year}{2006}).

\bibitem{Witzel:MultPulse}
\bibinfo{author}{Witzel, W.~M.} \& \bibinfo{author}{{Das Sarma}, S.}
\newblock \bibinfo{title}{Multiple-pulse coherence enhancement of solid state
  spin qubits}.
\newblock \emph{\bibinfo{journal}{Phys. Rev. Lett.}}
  \textbf{\bibinfo{volume}{98}}, \bibinfo{pages}{077601}
  (\bibinfo{year}{2007}).

\bibitem{Lee2008}
\bibinfo{author}{Lee, B.}, \bibinfo{author}{Witzel, W.~M.} \&
  \bibinfo{author}{Sarma, S.~D.}
\newblock \bibinfo{title}{Universal pulse sequence to minimize spin dephasing
  in the central spin decoherence problem}.
\newblock \emph{\bibinfo{journal}{Phys. Rev. Lett.}}
  \textbf{\bibinfo{volume}{100}}, \bibinfo{pages}{160505}
  (\bibinfo{year}{2008}).

\bibitem{Meiboom1958}
\bibinfo{author}{Meiboom, S.} \& \bibinfo{author}{Gill, D.}
\newblock \bibinfo{title}{Modified spin-echo method for measuring nuclear
  relaxation times}.
\newblock \emph{\bibinfo{journal}{Rev. Sci. Inst.}}
  \textbf{\bibinfo{volume}{29}}, \bibinfo{pages}{688} (\bibinfo{year}{1958}).

\bibitem{Khodjasteh:CDD}
\bibinfo{author}{Khodjasteh, K.} \& \bibinfo{author}{Lidar, D.~A.}
\newblock \bibinfo{title}{Performance of deterministic dynamical decoupling
  schemes: Concatenated and periodic pulse sequences}.
\newblock \emph{\bibinfo{journal}{Phys.\ Rev.\ A}}
  \textbf{\bibinfo{volume}{75}}, \bibinfo{pages}{062310}
  (\bibinfo{year}{2007}).

\bibitem{Uhrig2007}
\bibinfo{author}{Uhrig, G.~S.}
\newblock \bibinfo{title}{Keeping a quantum bit alive by optimizing $\pi$-pulse
  sequences}.
\newblock \emph{\bibinfo{journal}{Phys. Rev. Lett.}}
  \textbf{\bibinfo{volume}{98}}, \bibinfo{pages}{100504}
  (\bibinfo{year}{2007}).

\bibitem{Biercuk:ODD}
\bibinfo{author}{Biercuk, M.~J.} \emph{et~al.}
\newblock \bibinfo{title}{{Optimized dynamical decoupling in a model quantum
  memory}}.
\newblock \emph{\bibinfo{journal}{{Nature}}} \textbf{\bibinfo{volume}{{458}}},
  \bibinfo{pages}{{996--1000}} (\bibinfo{year}{{2009}}).

\bibitem{Du:ODD}
\bibinfo{author}{Du, J.} \emph{et~al.}
\newblock \bibinfo{title}{Preserving electron spin coherence in solids by
  optimal dynamical decoupling}.
\newblock \emph{\bibinfo{journal}{Nature}} \textbf{\bibinfo{volume}{461}},
  \bibinfo{pages}{1265--1268} (\bibinfo{year}{2009}).

\bibitem{Levy2002}
\bibinfo{author}{Levy, J.}
\newblock \bibinfo{title}{Universal quantum computation with spin-1/2 pairs and
  heisenberg exchange}.
\newblock \emph{\bibinfo{journal}{Phys.\ Rev.\ Lett.}}
  \textbf{\bibinfo{volume}{89}}, \bibinfo{pages}{147902}
  (\bibinfo{year}{2002}).

\bibitem{Field1993}
\bibinfo{author}{Field, M.} \emph{et~al.}
\newblock \bibinfo{title}{Measurements of coulomb blockade with a noninvasive
  voltage probe}.
\newblock \emph{\bibinfo{journal}{Phys.\ Rev.\ Lett.}}
  \textbf{\bibinfo{volume}{70}}, \bibinfo{pages}{1311} (\bibinfo{year}{1993}).

\bibitem{Neder2010}
\bibinfo{author}{Neder, I.}, \bibinfo{author}{Rudner, M.},
  \bibinfo{author}{Bluhm, H.} \& \bibinfo{author}{Yacoby, A.}
\newblock \bibinfo{title}{Semi-classical model for dephasing of an electron
  spin qubit coupled to a nuclear spin bath}.
\newblock \bibinfo{note}{(to be published)}.

\bibitem{Sundfors:NMR}
\bibinfo{author}{Sundfors, R.~K.}
\newblock \bibinfo{title}{Exchange and quadrupole broadening of nuclear
  acoustic resonance line shapes in the {III-V} semiconductors}.
\newblock \emph{\bibinfo{journal}{Phys. Rev.}} \textbf{\bibinfo{volume}{185}},
  \bibinfo{pages}{458--472} (\bibinfo{year}{1969}).

\bibitem{Taylor2007}
\bibinfo{author}{Taylor, J.~M.} \emph{et~al.}
\newblock \bibinfo{title}{Relaxation, dephasing, and quantum control of
  electron spins in double quantum dots}.
\newblock \emph{\bibinfo{journal}{Phys.\ Rev.\ B.}}
  \textbf{\bibinfo{volume}{76}}, \bibinfo{pages}{035315}
  (\bibinfo{year}{2007}).

\bibitem{Hester:defect}
\bibinfo{author}{Hester, R.~K.}, \bibinfo{author}{Sher, A.},
  \bibinfo{author}{Soest, J.~F.} \& \bibinfo{author}{Weisz, G.}
\newblock \bibinfo{title}{Nuclear-magnetic-resonance detection of charge
  defects in gallium arsenide}.
\newblock \emph{\bibinfo{journal}{Phys. Rev. B}} \textbf{\bibinfo{volume}{10}},
  \bibinfo{pages}{4262--4273} (\bibinfo{year}{1974}).

\bibitem{Bluhm2009}
\bibinfo{author}{Bluhm, H.}, \bibinfo{author}{Foletti, S.},
  \bibinfo{author}{Mahalu, D.}, \bibinfo{author}{Umansky, V.} \&
  \bibinfo{author}{Yacoby, A.}
\newblock \bibinfo{title}{Enhancing the coherence of spin quibts by narrowing
  the nuclear spin bath with a quantum feedback loop} (\bibinfo{year}{2010}).
\newblock \bibinfo{note}{(submitted to PRL)}, \eprint{arXiv:1003.4031}.

\end{thebibliography}
\begin{addendum}
\item[Acknowledgments] We thank D. J. Reilly for advice on
implementing the RF readout system and Jero Maze for discussions.
We acknowledge funding from ARO/IARPA, the Department
of Defense and the National Science Foundation under award number
0653336. I.N. and M.R. were supported by NSF grant DMR-0906475.
This work was performed in part at the Center for Nanoscale
Systems (CNS), a member of the National Nanotechnology Infrastructure
Network (NNIN), which is supported by the National Science Foundation
under NSF award no. ECS-0335765.  
\item[Author Contributions] 
e-beam lithography and MBE growth were carried out by D.M. and V.U.,
respectively.  H.B., S.F. and A.Y. fabricated the sample, planned and
executed the experiment and analyzed the data.  I.N., M.R. and H.B. and A.Y.
developed the theoretical model.  H.B., S.F. and A.Y., I.N. and
M.R. wrote the paper.

 \item[Correspondence] Correspondence and requests for materials 
should be addressed to  A.Y. (email:  yacoby@physics.harvard.edu).
\end{addendum}

\end{document}